%% file: main.tex
\documentclass[a4paper,11pt]{article}
\usepackage[utf8]{inputenc} 
\usepackage[spanish,english]{babel} 
\usepackage{amsmath,amsfonts,amssymb}
\usepackage{dcolumn}
\usepackage{booktabs}
\usepackage{txfonts} 
\usepackage{authblk} 
\usepackage{graphicx} 
\usepackage{float} 
\usepackage{multirow} 
\usepackage{makecell} 

\usepackage{caption} 
\usepackage{hyperref} 
\usepackage[numbers,comma,sort&compress,square]{natbib}
\makeatletter 
\def\NAT@def@citea{\def\@citea{\NAT@separator}}
\makeatother
\usepackage{multibib}
\newcites{supp}{Supplementary References}
\usepackage[table,xcdraw]{xcolor} 
\usepackage[margin=1.5cm]{geometry} 
\usepackage{lineno} 
\usepackage[misc]{ifsym} 
\usepackage{lipsum} 
\usepackage{sidecap} 

\usepackage{comment}
\usepackage[export]{adjustbox}
\usepackage{overpic}
\usepackage{multirow}
\usepackage{colortbl}
\usepackage{subcaption}
\usepackage{nameref}
\usepackage[capitalize,noabbrev]{cleveref}

\usepackage{cancel}
\usepackage[symbols,nogroupskip,nonumberlist,nopostdot=false,abbreviations]
{glossaries-extra}
\usepackage{array}
\usepackage{colortbl}

\usepackage{soul}

\input{glossary}
\usepackage{tcolorbox}
\usepackage{siunitx}

\usepackage{geometry}
\usepackage{tikz}
\usetikzlibrary{shapes.geometric, arrows, positioning}

\usepackage{media9}

\tikzstyle{startstop} = [rectangle, rounded corners, minimum width=3cm, minimum height=1cm, text centered, draw=black, fill=red!30]
\tikzstyle{process}   = [rectangle, minimum width=3cm, minimum height=1cm, text centered, draw=black, fill=blue!20]
\tikzstyle{decision}  = [diamond, aspect=2, text centered, draw=black, fill=green!30]
\tikzstyle{arrow}     = [thick,->,>=stealth]

\usepackage{algorithm}
\usepackage{algpseudocode}
\usepackage{siunitx}
\usepackage{enumerate}
\usepackage{enumitem}
\usepackage{caption}
\usepackage{tcolorbox}

\newcounter{myframecounter}
\newenvironment{myframe}[1]{%
    \refstepcounter{myframecounter}%
    \begin{tcolorbox}[colback=white,colframe=black,title=\textbf{Box \themyframecounter.} #1]\par\medskip
}{%
    \end{tcolorbox}
}
\crefname{myframecounter}{Box}{Boxes}

\newlist{myenumerate}{enumerate}{1}
\setlist[myenumerate]{label=\textbf{\arabic*.}}

\newlist{myitemize}{itemize}{1}
\setlist[myitemize]{label=\textbf{$\bullet$}}

\newlist{mydescription}{description}{1}
\setlist[mydescription]{font=\normalfont\bfseries}

\makeatletter\renewcommand{\@biblabel}[1]{#1.}\makeatother 
\addto{\captionsenglish}{}
\hypersetup{ 
  colorlinks = true,
  citecolor=  blue,
  linkcolor = {blue},
  filecolor = cyan, 
  urlcolor = blue
}
\DeclareCaptionLabelSeparator{vbar}{~\textbar~}
\DeclareCaptionLabelSeparator{dot}{.~}
\definecolor{lightgray}{gray}{0.9}
\newcommand{\figref}[2][]{%
  \hyperref[{#2}]{%
    Fig.~\ref*{#2}%
    \ifx\\#1\\%
    \else
      #1%
    \fi
  }%
}
\newcommand{\tableref}[1]{%
  \hyperref[{#1}]{%
   Table~\ref*{#1}%
  }%
}
\newcommand{\methodsref}[1]{%
  \hyperref[{methods:#1}]{%
   Methods:~\nameref*{methods:#1}%
  }%
}
\makeatletter 
\renewcommand{\maketitle}{\bgroup\setlength{\parindent}{0pt}
\begin{flushleft}
  \textbf{\LARGE \@title}
  
  \bigskip
  
  \@author
\end{flushleft}\egroup
}
\makeatother
\renewenvironment{abstract} 
 {\par\noindent\textbf{\abstractname~\textbar}\ \ignorespaces}
 {\par\medskip}

\makeatletter
\newcommand{\setcurrentname}[1]{\def\@currentlabelname{#1}}
\makeatother

\title{\textit{Structural causal influence} (\textit{SCI}) captures the forces of social inequality in models of infectious disease}

\author[1,2]{Sudam Surasinghe}
\author[1]{Swathi Nachiar Manivannan}
\author[3,4,5]{Samuel V. Scarpino}
\author[6]{Lorin Crawford}
\author[1,2,5 \Letter]{C. Brandon Ogbunugafor}
\affil[1]{Department of Ecology \& Evolutionary Biology,
Yale University, New Haven, CT, 06511, USA}
\affil[2]{Public Health Modeling Unit, Yale School of Public Health, New Haven, CT, 06511 USA}
\affil[3]{Institute for Experiential AI, Northeastern University, Boston, Massachusetts, 02115, USA}
\affil[4]{Network Science Institute, Northeastern University, Boston, Massachusetts, 02115, USA}
\affil[5]{Santa Fe Institute, Santa Fe, NM, 87501, USA}
\affil[6]{Microsoft Research, Cambridge, MA, 02142, USA}
\affil[$\textrm{\Letter}$]{CBO: brandon.ogbunu@yale.edu}
\date{}

\begin{document}

\maketitle
\bigskip\bigskip
\begin{abstract}
Mathematical modeling has played a central role in understanding how infectious disease transmission manifests in populations. These models have demonstrated the importance of key community-level factors in structuring epidemic risk, and are now routinely used in public health for decision support. One barrier to their broader utility is that the existing canon does not often accommodate social inequalities as distinct formal drivers of variability in transmission dynamics. Given decades of evidence supporting the organizational effects of inequalities in structuring society more generally, and infectious disease risk more specifically, addressing this modeling gap is of critical importance. In this study, we build on previous efforts to integrate social forces into computational epidemiology by introducing a metric, the \textit{structural causal influence} (\textit{SCI}). The \textit{SCI} uses causal analysis to provide a measure of the relative vulnerability of sub-communities within a susceptible population, shaped by differences in characteristics such as access to therapy, exposure to disease, and other determinants driven by social forces. We develop our metric in a simple case and apply it to a context of public health importance: Hepatitis C virus in a population of persons who inject drugs. In addition, we demonstrate the flexibility of the \textit{SCI} using an agent-based model of an infectious disease. Our use of the \textit{SCI} reveals that, under specific parameters in a multi-community model, the "less vulnerable" community may achieve a basic reproduction number below one, ensuring disease extinction. However, even minimal transmission between communities can increase this number, leading to sustained epidemics within both communities.

\vspace{1em}
\noindent\textbf{Keywords:} computational epidemiology, disease ecology, causal analysis, social determinants of health
.

\end{abstract}
\section{Introduction}

In recent decades, scholars have questioned several classical assumptions of computational epidemiology and disease ecology \cite{roberts2015,brauer2017}. For example, epidemiological models have long relied on reproduction numbers to characterize disease transmission dynamics. However, their interpretation is often based on the assumption of a homogeneous population, making it less reliable when applied to heterogeneous populations with varying social, demographic, and behavioral factors.

In recent times, epidemiological modeling has evolved to incorporate important features beyond classical formulations, especially in addressing population heterogeneity through subgroups with differing within- and between-group contact rates~\cite{haskey1957stochastic, hethcote1987epidemiological, longini1978optimization, brauer2008epidemic, danon2011networks}. In modeling sexually transmitted infections, Lajmanovich and Yorke~\cite{lajmanovich1976mathematical} introduced stratification by contact rates and susceptibility, which was expanded upon by Hethcote and Yorke~\cite{hethcote1978dynamics}, who emphasized the role of high-prevalence \emph{core groups} in transmission and control. To move beyond limitations of classical epidemic summaries like the basic reproduction number $\mathcal{R}_0$, researchers developed refined metrics such as the \textit{Target} or \textit{Type Reproduction Number}(TRN)  \cite{roberts2003new, heesterbeek2007type, shuai2013extending, roberts2003new, heesterbeek2007type, inaba2013definition, lewis2019general, wang2020target}, which quantify subpopulation-specific dynamics using projection and selection operators on the next-generation matrix~\cite{roberts2003new}. The TRN is part of a broader literature challenging the sufficiency of basic assumptions in epidemiology by accounting for heterogeneous secondary contacts~\cite{hebertdufresne2020r0}, and parallels developments such as the effective reproduction number $\mathcal{R}_\text{t}$~\cite{gostic2020rt} and network- or simulation-based models that better capture real-world complexities~\cite{keelingeames2005, garnerhamilton2011, iranzo2021, crepey2022}. These also include the importance of host behavior~\cite{bansal2007individual}, and phenomena like complex contagion~\cite{guilbeault2018, hebertdufresne2020contagions, stonge2024}, adaptive behavior~\cite{scarpino2016effect}, and superspreading~\cite{lloydsmith2005}.

In human infectious disease research, a century of epidemiological study has established strong links between health outcomes and social forces such as socioeconomic status \cite{marmot2005social}. As complexity scientists Melanie Moses and Kathy Powers succinctly noted: ``Well-mixed models do not protect the vulnerable in segregated societies'' \cite{powersmitchelltransmission}. In response, scholars have called for the incorporation of social determinants into the study of infectious diseases \cite{bedson2021, zelner2022, galanis2021incorporating, glennon2021challenges}, and have shown that targeting disadvantaged populations can improve outcomes for diseases such as tuberculosis \cite{andrews2015epidemiological}. A growing body of work emphasizes the role of socioeconomic variables in shaping infectious disease disparities \cite{wang2024modifiable, manna2024importance, frey2024inequalities, silva2024income}, and demonstrates how integrating these factors into computational models—via generalized contact matrices or high-resolution mixing data—can improve model fidelity and relevance \cite{tizzoni2022addressing, mistry2021inferring, volz2011effects, manna2024generalized}. Despite these advances, most models still fall short of quantifying the causal relationship between structural inequality and disease outcomes, and lack discrete metrics to measure the impact of this inequality across model types. 

Traditional epidemiological metrics, such as the population attributable fraction (PAF) introduced by Levin in 1953 \cite{ml1953occurrence}, have been instrumental in estimating the burden of disease that can be attributed to specific risk factors. Levin's formula \cite{ml1953occurrence} utilizes the prevalence of exposure and the relative risk to assess the potential impact of removing a risk factor from the population. However, this approach assumes independence among individuals and does not consider the complex dynamics of infectious disease transmission, such as interactions between individuals and varying susceptibility across subpopulations. Recognizing these limitations, subsequent models \cite{brooks2017defining} have introduced parameters to account for the proportion of contacts occurring within a group, thereby incorporating aspects of interaction between individuals. 

Building upon these advancements, we introduce \gls{sciAb}, a framework rooted in causal inference that systematically quantifies the impact of heterogeneity on disease transmissions. It provides a metric that captures the role of social forces in any model where basic reproductive ratios can be computed. We define \gls{sciAb} based on the estimation of basic reproduction numbers for sub-communities that compose a greater susceptible population. We first demonstrate the theoretical formulation of the \gls{sciAb} using classical \gls{ode}-based epidemiological models, where stratification is explicitly introduced based on social inequalities. In particular, we focus on a more complex case involving the \gls{hcv} in a population of persons who inject drugs (PWID). We then provide a more general simulation framework based on \gls{ABM} and present data-driven estimation techniques for the \gls{sciAb}. 

We discuss the \gls{sciAb} with respect to other approaches that examine heterogeneity in epidemiological models. For example: Unlike TRN, our approach does not rely solely on the next-generation matrix, but is capable of incorporating data-driven methodologies for estimation, facilitating application to real epidemiological data and simulated data from complex models. Furthermore, unlike the core-group concept (that qualitatively highlights the role of high-prevalence sub-populations), the \gls{sciAb} provides a rigorous, quantitative measure of influence attributable to structural disparities—such as differential exposure, unequal access to healthcare, and behavioral constraints. Because of this, we offer that the \textit{SCI} can be considered to exist in the tradition of innovations that add richness to epidemiological models and supports broader efforts to integrate social determinants into both the technical realms of epidemiology and disease ecology.

\section{Materials and Methods} 
In this section, we introduce the \textit{structural causal influence} and argue that it functions as a metric to gauge how the social forces that drive inequalities between susceptible populations influence disease dynamics. Before discussing the applications of the \gls{sciAb}, we first outline its formulation and interpretation. We then demonstrate the use of the metric through applications in both compartmental modeling and the \gls{ABM} framework. For the reader's convenience, we summarize the structure of the article, highlighting the key sections that present results from the application of \gls{sciAb} as well as the underlying design concepts of the models.   Note that the parenthetical ``SD" describes models that include the role of social forces in disease dynamics, through the \textit{SCI}. In the main text below, we summarize the three different modeling scenarios that we used to illustrate the \gls{sciAb} framework. We note that the Supporting Information contains a more detailed treatment of computational methods for all of the examples.

\begin{enumerate}  
    \item An SIR model that incorporates social determinants into the classical Susceptible– Infected–Recovered (SIR) epidemiological framework is presented in the Supporting Information (see the ``SIR(SD) Model: Extending the SIR model to incorporate social inequalities" section). Results based on specific hyperparameters are outlined in the ``Analysis of \gls{sciAb} in the SIR(SD) model with hyperparameters driven by social inequalities" section of the Supporting Information. The analysis of the measures capturing social inequalities within the SIR model is provided in the ``Analysis of social inequality measures in the SIR(SD) model across the full parameter space" section of the Supporting Information.
    \item A model of \glsdesc{hcv} that characterizes the dynamics of an outbreak in a population consisting of two distinct social communities of persons who inject drugs (PWID).The results are summarized in the ``\nameref{sec:HCVsdRes}'' section, with a detailed discussion and computational specifications of the model provided in the ``HCV(SD) model: hepatitis C virus model affected by social determinants'' section of the Supporting Information.
  \item   An \glsdesc{ABM} example, offering a complementary perspective on the applicability of the metric. The model simulates agents/individuals moving in a continuous 2D space, each of whom can transition through Susceptible, Exposed, Infectious, and Recovered (S-E-I-R) states. Disease transmission is modeled on the basis of spatial proximity, with the likelihood of infection depending on the local density of infectious individuals. The results are summarized in the ``\nameref{sec:RestABM}" section, with a comprehensive discussion of the model in the ``Agent-based model incorporating health inequalities" section of the Supporting Information.

\end{enumerate}  

Evaluating \gls{sciAb} using compartmental models provides analytical insight into the measure, while \gls{ABM} inherently accommodates heterogeneity in a direct manner by allowing different parameter values to be assigned to individual agents. Consequently, incorporating social inequalities into \gls{ABM} is relatively easy to implement. In contrast, compartmental models require additional compartments and parameters to account for stratification based on social inequalities.

\subsection{\textit{The structural causal influence} of a social community}\setcurrentname{\textit{Structural causal influence} of a social community}\label{sec:mesSDOH}

Structural causal models, as introduced in \cite{pearl2000causality,pearl2012calculus}, can be used to measure the effect of social community heterogeneity on the dynamics of the disease. We first use a deterministic version of structural causal models for \glspl{ode} as discussed in a prior study \cite{mooij2013ordinary}. \textit{Interventions} in the compartmental disease models with social inequalities can be implemented in different ways. We utilize ``perfect interventions'' \cite{mooij2013ordinary} since our goal is to determine the ``individual" influence of each social community on the overall dynamics of the system. In the context of the SIR(SD) model, \textit{perfect interventions} mean that we will set the value of the infectious population of the considered community to be zero and assume that the intervention is activated from time $t=0$ to $t\rightarrow \infty$. This means that the considered infectious population is absent and remains unchanged over time. We follow the notion described in \cite{mooij2013ordinary}, inspired by the do-operator introduced by \cite{pearl2000causality}. We will denote this type of intervention as $\text{do}(\Gamma)$, where $\Gamma$ represents the conditions required for \textit{perfect interventions}.
Using the change in $\mathcal{R}_0$ due to social inequalities, the \gls{sciAb} of social community $i$ on disease dynamics can be defined as follows:
\begin{equation}\label{eq:SCIcomp}
    \mathcal{C}_i=\frac{\mathcal{R}_0^{\text{SD}}-\mathcal{R}_0^{\text{do}(\Gamma_i)}}{\mathcal{R}_0^{\text{SD}}},
\end{equation}

where $\mathcal{R}_0^{\text{SD}}$ is the basic reproduction number for the compartmental model incorporating social inequalities, and $\mathcal{R}_0^{\text{do}(\Gamma_i)}$ denotes the basic reproduction number for the same model with intervention for community $i$.  
For any general epidemiological model, the computational steps presented in \cref{fram:SCI} can be used to calculate the \gls{sciAb}.

\begin{myframe}{\small Computational Framework for \textit{SCI}}\label{fram:SCI}
    \begin{myenumerate}
        \item \textbf{Full model with social inequalities:} Develop an extended epidemiological model incorporating multiple social communities to account for parameters influenced by social inequalities.
        \begin{myitemize}
        \item Calculate the basic reproduction number, $\mathcal{R}_0^{\text{SD}}$, for the full model.
        \end{myitemize}
        \item \textbf{Reduced model with \textit{perfect interventions} applied to community $\boldsymbol{i}$:} Assuming that \textit{perfect interventions} are applied to community $i$ (implying no infection or transmission from community $i$ is possible), the full model is reduced to the community $i$ \textit{perfect interventions} epidemiological model.  
        \begin{myitemize}
        \item Calculate the basic reproduction number, $\mathcal{R}_0^{\text{do}(\Gamma_i)}$, for the reduced model with \textit{perfect interventions} applied to community $i$.
        \end{myitemize}
   
    \item \textbf{Compute the \glsdesc{sciAb}, $\boldsymbol{\mathcal{C}_i}$:} Using the basic reproduction numbers calculated for the full model (in Step 1) and the reduced model (in Step 2), compute the \gls{sciAb} for community $i$ using \cref{eq:SCIcomp}: 
\begin{equation*}
    \mathcal{C}_i=\frac{\mathcal{R}_0^{\text{SD}}-\mathcal{R}_0^{\text{do}(\Gamma_i)}}{\mathcal{R}_0^{\text{SD}}}.
\end{equation*}  

\textbf{Note:} The value of $\mathcal{R}_0^{\text{do}(\Gamma_i)}$ is upper bound by $\mathcal{R}_0^{\text{SD}}$ if the \textit{perfect intervention} in community $i$ does not alter the parameters of the non-intervened community. However, such an intervention can introduce changes in other estimated parameters, which can result in scenarios where $\mathcal{R}_0^{\text{do}(\Gamma_i)} > \mathcal{R}_0^{\text{SD}}$. However, we may reasonably assume that these variations are sufficiently small, ensuring that $\mathcal{R}_0^{\text{do}(\Gamma_i)}$ remains close to $\mathcal{R}_0^{\text{SD}}$. Consequently, the bounds $-1 < \mathcal{C}_i \le 1$ hold for community $i$.

     \end{myenumerate}
\end{myframe}

\subsubsection{Notes on \glsdesc{sciAb} (\textit{SCI}, \texorpdfstring{$\boldsymbol{\mathcal{C}_i}$}{C_i})}
 To contextualize the structural causal influence (SCI) metric, we briefly compare it to the population attributable fraction (PAF) for infectious diseases. Brooks-Pollock and Danon~\cite{brooks2017defining} proposed a formulation for PAF in terms of the basic reproduction number, which quantifies the proportionate reduction in transmission that would occur if a given risk group were removed. They demonstrate that this definition is mathematically equivalent to the traditional PAF under scenarios where the risk group affects only susceptibility.

In contrast, the SCI metric introduced here is defined using Pearl’s \texttt{do}-calculus and is grounded in the framework of structural causal models. Rather than assuming uniform effects or static population structure, SCI formalizes the causal influence of a specific sub-community by comparing system-level transmission dynamics with and without that group’s contribution. In this way, SCI generalizes the intuition behind Brooks-Pollock’s formulation to stratified, dynamic models with heterogeneous mixing and explicit social determinants. Given their resonance, however, one can consider the \gls{sciAb} to be a modification of the PAF, and in this study, applied to differences in infectious disease dynamics as a function of social forces. A numerical comparison of the proposed \gls{sciAb} and PAF measures within the SIR model framework is provided in Remark 1 of the Supporting Information.

The \gls{sciAb} offers a comparative analysis of disease dynamics within the full model, encompassing multiple social communities, and a model representing the effects of \textit{perfect interventions} within one specific community. It quantifies the disparity in secondary infections between the total population and the population under \textit{perfect interventions} for the designated social community. To standardize this measure, it is normalized by the total secondary infections within the population, resulting in a scale ranging from -1 to 1. A value closer to zero suggests a convergence between the dynamics of the full model and the \textit{perfect intervention} model for the specified community. In such cases, the influence of the considered community is relatively minimal, indicating an inability to stop disease transmission solely within that community. Conversely, a higher \textit{SCI} value indicates discrepancies between the dynamics of the full model and the \textit{perfect intervention} model, highlighting an influence exerted by that particular community. Specific features of the \textit{SCI} can be summarized as follows:
\begin{itemize}
\item The \textit{SCI} value $\mathcal{C}_i$ for a given social community $i$ lies within the range $-1 < \mathcal{C}_i \leq 1$, where $\mathcal{C}_i$ measures the influence of community $i$ on disease dynamics.
\item In scenarios with only two social communities $i$ and $j$, the basic reproduction number for \textit{perfect interventions} within community $i$ is equal to the basic reproduction number for an isolated community $j$, devoid of inter-community contacts.
\item For two social communities $1$ and $2$, the \textit{SCI} value of $\mathcal{C}_1\approx 0$ indicates that community $2$ exhibits the highest basic reproduction number in the absence of inter-community contacts.
\item In the context of two social communities $1$ and $2$, the \textit{SCI} value of $\mathcal{C}_1=1$ signifies that community $2$ does not experience any secondary infections in the absence of inter-community contacts.

\item In the context of two social communities, $1$ and $2$, a negative \textit{SCI} value for community 1 ($\mathcal{C}_1 < 0$) suggests that the epidemic curve of the total population is significantly negatively affected by community 1. In this case, the isolated community 2 has a higher basic reproduction number than the full model. Example scenarios include situations where the disease dies out in isolated community 1, while an epidemic persists in isolated community 2. As a result, the epidemic curve for the total population, when compared to that of isolated community 2, exhibits a substantial downward shift (along the y-axis). This shift is due to the influence of community 1.
\end{itemize}

We first present general considerations for incorporating social forces into compartmental models prior to further analysis.
\paragraph*{Incorporating social forces in epidemiological compartmental models:} 
In classical models, host populations are divided into different compartments, where homogeneity is assumed between individuals within the same compartment, and the movement of individuals between compartments is parameterized by a system of \glspl{ode}. Although useful in helping epidemiologists analyze the dynamics of infectious diseases, the underlying assumptions of the parameter value distributions and homogeneity are not representative of real-world populations. In particular, one can consider how health inequalities can shape parameter values \cite{blumenshine2008pandemic, diderichsen2001social}:
\begin{itemize}
    \item 
    \textit{Differences in exposure} \\
    Examples: Crowding in households \cite{baker2013infectious,cardoso2004crowding, rader2020}, medical facilities, public transportation, and/or occupational factors such as the inability to work from home \cite{zhang2021covid} or dependence on childcare outside the home \cite{holmes1996child}.
    \item \textit{Differences in susceptibility} \\
    Examples: Host factors, such as pre-existing immunity, age, other underlying diseases or conditions \cite{marais2013tuberculosis}, smoking and/or environmental tobacco smoke \cite{huttunen2011smoking}, nutritional status \cite{brown2003diarrhea,rodriguez2011malnutrition}, stress, and/or vaccination status, where there may be differences in vaccine acceptance, uptake, and/or access \cite{kolobova2022vaccine}.
    \item \textit{Differences in timely and effective treatment} \\
    Examples: Access to outpatient and inpatient medical care \cite{kronman2022historical}, care-seeking attitudes and behavior \cite{wamala2007perceived}, financial obstacles (including lack of adequate insurance coverage) \cite{santoli2004insurance}, and logistical obstacles, such as transportation, language \cite{guirgis2012barriers}, quality of care, and/or availability of treatments.
\end{itemize}
Within compartmental models, parameters, variations in susceptibility and exposure often impact the transmission rate, while differences in timely and effective treatment may influence the mean infectious period.

\paragraph*{Definitions:} We use the term ``social forces'' as an umbrella term to describe the characteristics of a social structure that can foster inequalities with respect to disease burden, or in terms of vulnerability to harm. It encompasses concepts of interest to the public health sector, such as the social determinants of health, or health inequality. We utilize the more general term, as our methods also apply ``social forces" that may persist in non-human host-pathogen systems. We also refer to sub-populations in our models as ``community 1" and ``community 2", since our models consider the specific scenario where there are two distinct social sub-populations. 

\section{Results} 
\subsection{Applying \textit{SCI} to the HCV(SD) compartmental model}\setcurrentname{Applying \textit{SCI} to a HCV(SD) compartmental model}\label{sec:HCVsdRes}

\subsubsection{Description of the HCV(SD) model}
 In the Supporting Information, we introduce methods for modeling the influence of social forces using an iteration on a classical standard SIR model. Here, we examine a more complex example of disease ecology, one that more directly invokes the role of social forces: \glsdesc{hcv} in a population of persons who inject drugs (PWID). \gls{hcv} has long been understood as a disease where social inequalities of various sorts influence disparate risk, treatment, and outcomes \cite{lelutiu2009meta,ford2017neighborhood,vaz2025socioeconomic}. 

\begin{figure}[ht!]
    \centering
\begin{subfigure}{0.45\textwidth}
        \centering
       \begin{overpic}[width=1\linewidth]{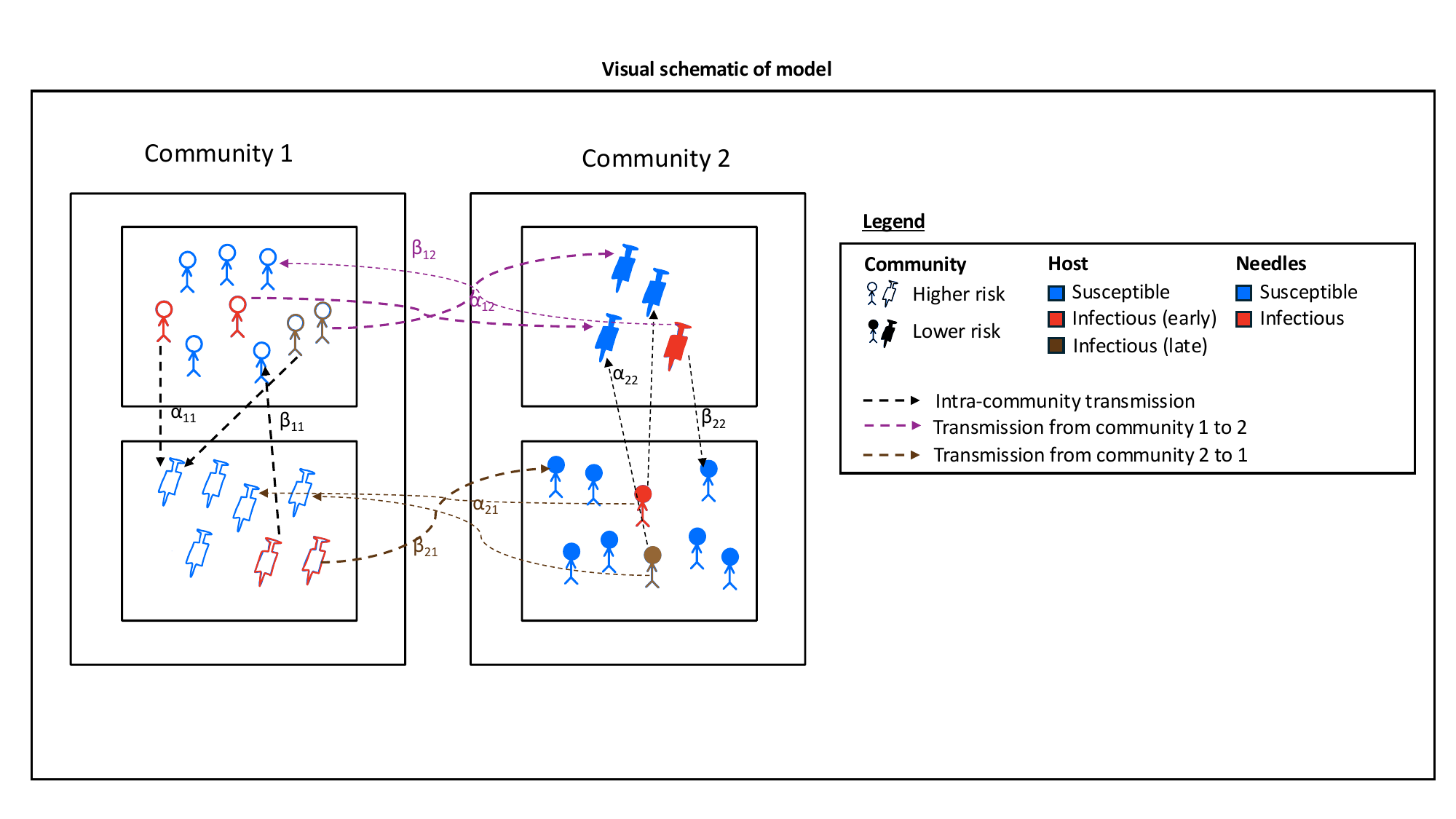}
        \put(-5,50){\textbf{(a)}}
    \end{overpic}
    \caption*{}
    \end{subfigure}
        \begin{subfigure}{0.45\textwidth}
        \centering
    \begin{overpic}[width=1\linewidth, valign=t]{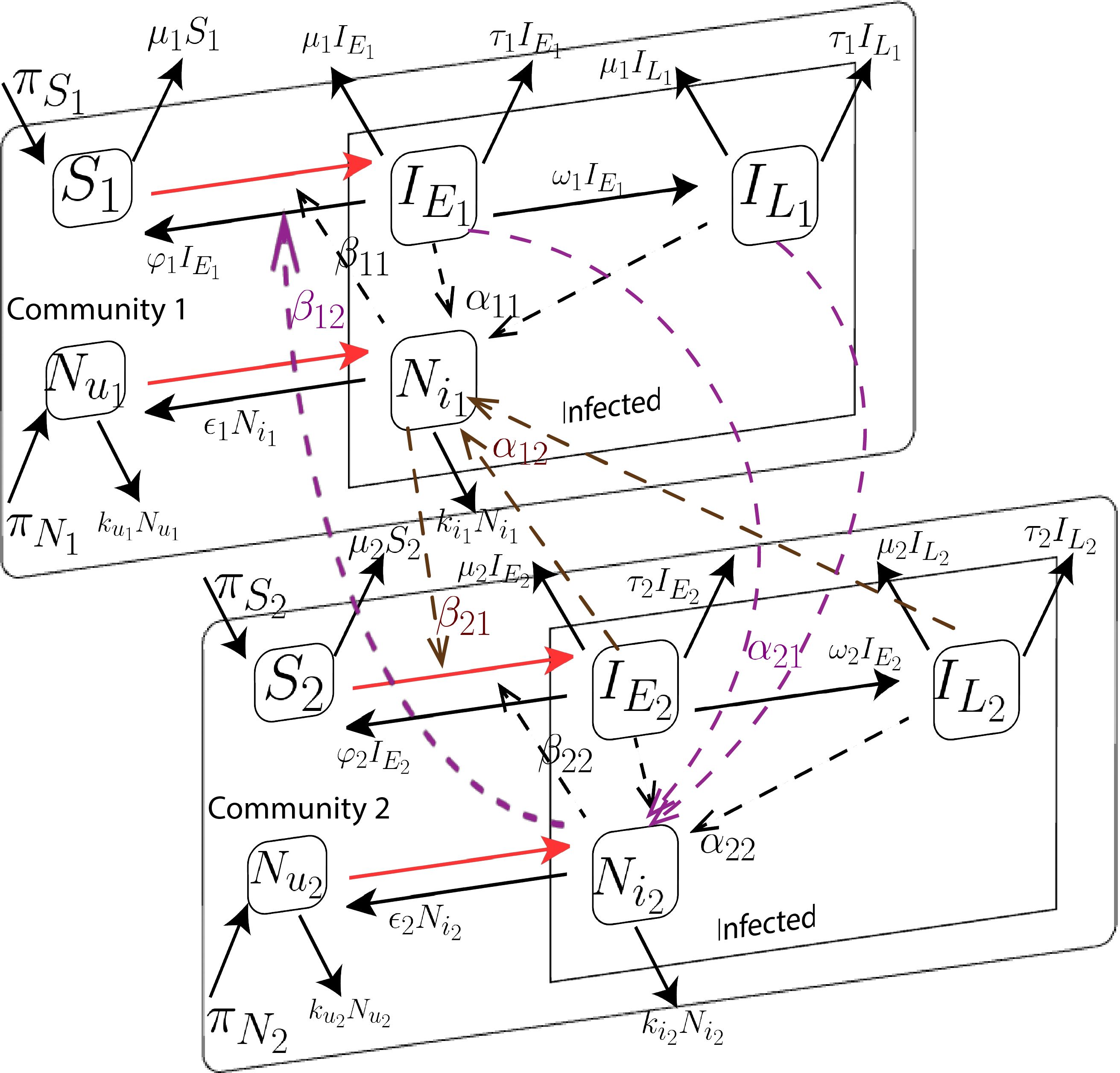}
        \put(-5,95){\textbf{(b)}}
    \end{overpic}
    \caption*{}
    \end{subfigure}
    \caption{\textbf{HCV mathematical model:} (a) A simplified schematic of a compartmental epidemiological model for \gls{hcv} with two socially distinct PWID communities, referred to as the HCV(SD) model. The schematic depicts how transmission is modeled between hosts and injection paraphernalia within and between the two PWID communities. In this case, community 1 is assumed to be the community of higher risk. (b)  Mathematical compartmental depiction of the HCV(SD) model. The details regarding the parameters and the system of equations underlying the model is detailed in the ``HCV(SD) Model: Hepatitis C Virus model affected by social determinants'' section of the Supporting Information. The parameter $\pi_S$ denotes the birth rate of susceptible individuals in a given community, while $\varphi$ represents the daily fractional self-clearance rate for the same group. The transition from early to late-stage infection is captured by $\omega$, and $\tau$ indicates the rate at which individuals in the community enter treatment. The parameter $\pi_N$ reflects the birth rate of uninfected needles circulating in the community, whereas $\epsilon$ accounts for the clearance (or decay) rate of \gls{hcv} infection in needles. The discard rates of uninfected and infected needles in the community are represented by $k_u$ and $k_i$, respectively. The parameter $\alpha_{jl}$ denotes the injection rate in community $l$ by needles circulating in community $j$, multiplied by the probability that the needle is infected, and $\beta_{jl}$ denotes the injection rate in community $j$ by needles circulating in community $l$, multiplied by the rate at which a host individual in community $j$ can become infected. Finally, $\mu$ denotes the removal rate from the community. Subscripts in the parameter notation (e.g., $\pi_{S_1}$, $\alpha_{12}$) are used to specify the referred community or source/target community pair. The complexity of the graph highlights the central challenge in modeling scenarios with population sub-structure: multiple forces are interacting within multiple populations. The methods and metrics proposed allow one to deconstruct this complexity. The model is adapted from a previous study of HCV dynamics in a single population \cite{miller-dickson_hepatitis_2019}.}
    \label{fig:summaryModel}
\end{figure}

The HCV(SD) model that we will employ is an extension of the waterborne, abiotic, and indirectly transmitted (W.A.I.T.) modeling framework used in several studies \cite{miller2019waterborne, miller-dickson_hepatitis_2019, meszaros2020direct, ogbunugafor2020variation}. An illustration of the HCV(SD) model is provided in \cref{fig:summaryModel}, and a comprehensive description can be found in the ``HCV(SD) Model: Hepatitis C Virus model affected by social determinants'' section of the Supporting Information. For the simulations in our study, we utilize a population size approximating $170,000$ individuals, based on estimations of the PWID population in New York City \cite{des1998declining, miller-dickson_hepatitis_2019}. This model accommodates the migration of people who inject drugs into the population. Within the framework of our HCV(SD) model (\cref{fig:summaryModel}), we assume two distinct PWID communities, each comprising an equal population size of $85,000$ individuals. As shown in \cref{fig:summaryModel} (b), the model consists of the following compartments: \( S_j \), \( I_{E_j} \), \( I_{L_j} \), \( N_{u_j} \), and \( N_{i_j} \), which denote the populations of susceptible individuals, early-stage infected individuals (representing acute \gls{hcv} infection), late-stage infected individuals (representing chronic \gls{hcv} infection), uninfected needles and infected needles, respectively, within the community \( j \). The population dynamics are evaluated by solving the system of ordinary differential equations presented in the .  Additionally, we explicitly differentiate all needles in circulation across PWID communities 1 and 2, ensuring a comprehensive representation of the transmission dynamics.

In these W.A.I.T. models, injection paraphernalia serves as the environmental reservoir for \gls{hcv}, with the sharing of this equipment constituting the primary transmission mode for new infections. Although injection paraphernalia includes various components, the parameters in these models are based on the use of needles and syringes as primary instruments of injection and sharing. Therefore, in this study, we use the term 'needle' as a synecdoche for the entire injection apparatus. It should be noted that, although \gls{hcv} can be transmitted sexually \cite{terrault2013sexual}, our study focuses on transmission through infected injection equipment. In addition, we assume the presence of two distinct needle populations associated with each community. In this model, needles do not cross communities, but people do. This structure enables us to examine the impact of social inequalities on communities of people, and the dynamics of \gls{hcv} transmission within and between communities. 

Using the \gls{sciAb} measure, we can attempt to understand the impact of social determinants on the dynamics of the disease by analyzing the control parameters. We will focus on three crucial parameters that can vary between communities. Firstly, the clearance rate of \gls{hcv} in needles ($\epsilon$) is expected to be influenced by social factors, such as the existence of harm reduction programs like safe injection sites (which replace potentially infected paraphernalia with uninfected paraphernalia \cite{van2017full}. Secondly, infected populations in different communities could exhibit differences in their self-clearance rates ($\varphi$) \cite{kaaberg2018incidence}. Lastly, the treatment rate ($\tau$) can be influenced by social determinants, as everyone does not have equal access to clinical care and therapy \cite{grebely2019elimination,beiser2019hepatitis}. In this analysis, we will consider relatively high values for these parameters in community 1 and lower values in community 2.

In addition, we will investigate the effect of population mixing. We examine four distinct mixing scenarios while keeping the intra-community transmission rates constant ($\beta_{11}=\beta_{22}=0.0360$ and $\alpha_{11}=\alpha_{22}=2$), but varying the inter-community transmission rates as follows:
\begin{enumerate}
\item Isolated ($\beta_{12}=\beta_{21}=0$ and $\alpha_{12}=\alpha_{21}=0$),
\item $1\%$ mixing ($\beta_{12}=\beta_{21}=0.00036$ and $\alpha_{12}=\alpha_{21}=0.02$),
\item $5\%$ mixing ($\beta_{12}=\beta_{21}=0.0018$ and $\alpha_{12}=\alpha_{21}=0.1$), or
\item Homogeneous mixing ($\beta_{12}=\beta_{21}=0.036$ and $\alpha_{12}=\alpha_{21}=2$).
\end{enumerate}

 Note that the $\beta$ parameters represent host transitions resulting from contact with infected needles, while the $\alpha$ parameters correspond to needle transitions mediated by infected hosts. For a more detailed description of the model, refer to the ``HCV(SD) Model: Hepatitis C Virus model affected by social determinants'' section of the Supporting Information.

\begin{figure}[!ht]
     \centering
\includegraphics[width=1\linewidth]{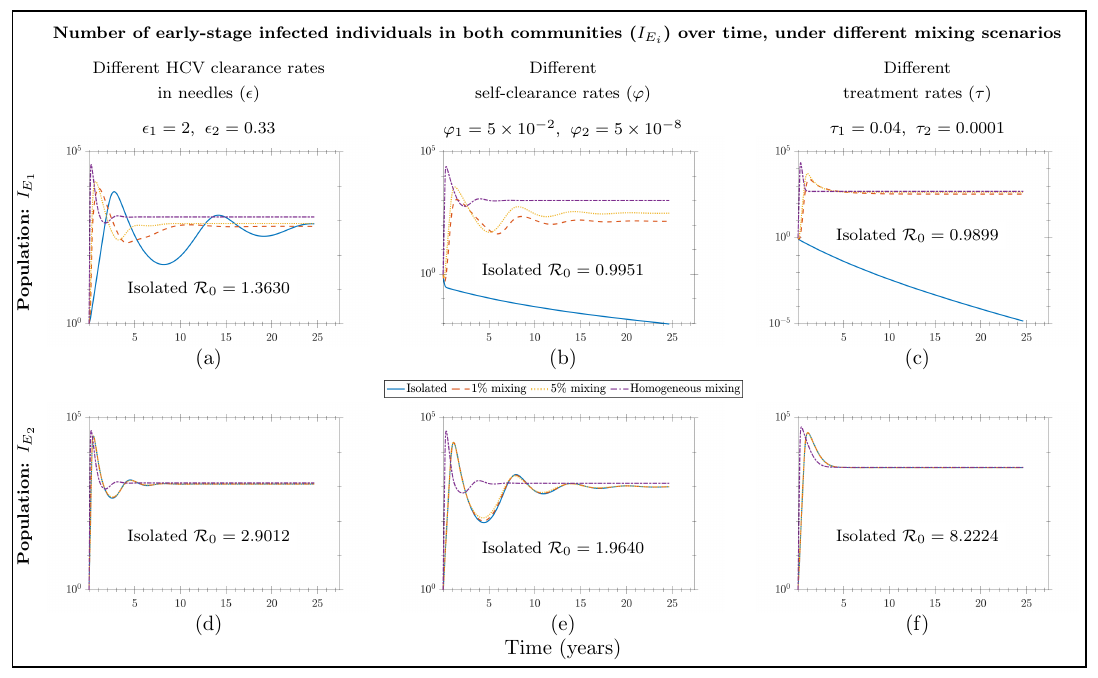}
\caption{\textbf{HCV(SD) mathematical model:} Population dynamics of early-stage infected individuals (in log scale) in two communities (community 1 (top) and community 2 (bottom)) with varying control parameters (\gls{hcv} clearance rate in needles ($\epsilon$) (left), self-clearance rate ($\varphi$) (middle), and treatment rate ($\tau$) (right)) under different mixing scenarios are illustrated. Each subplot depicts the early-stage infected population over time. Here, we have selected community 2 to be more vulnerable by adjusting the control parameters accordingly. Note that with the selected parameters, the less vulnerable community 1 may have a lower basic reproduction number when isolated (i.e., assuming \textit{perfect intervention} in community 2). Even with the specified self-clearance rate and treatment rate (see plots in the first row, second, and third columns), this basic reproduction number remains less than one, leading to the rapid die-out of the disease (see solid blue curves in those plots). However, even a small amount of inter-community transmission can trigger an epidemic in community 1 (see dashed red curves indicating $1\%$ mixing in those plots).}
      \label{fig:HCVResSummary}
\end{figure}

The isolated scenario can be regarded as the baseline model with no interaction between the two communities. Consequently, it can be conceptualized as comprising two sub-models, each representing its community dynamics. In contrast, homogeneous mixing results in a fully connected model, where individuals from either community have an equal chance of contact with anyone in the other community. Furthermore, low-mixing scenarios allow us to investigate the effects of minimal inter-community interactions on the dynamics of the entire community. This controlled exploration enables a nuanced understanding of how even subtle changes in mixing patterns can influence population disease dynamics. We use MATLAB's ode45 numerical solver \cite{shampine1997matlab} to estimate the population dynamics for each of the four mixing scenarios, setting the initial conditions of uninfected populations at their \gls{dfe} values ($S_1=S_2=85,000$ and $N_{u_1}=N_{u_2}=110,000$), with $I_{E_1}=I_{E_2}=N_{i_1}=N_{i_2}=1$, and $I_{L_1}=I_{L_2}=0$. \cref{fig:HCVResSummary} depicts the results of the different mixing scenarios, particularly showing the infected population in the early-stage (in log scale) for both communities.

\subsubsection{HCV(SD) model: the clearance rate for needles ($\boldsymbol{\epsilon}$)}
  In the dynamics of infectious diseases among PWID communities, the clearance rate of needles is an important feature. The \textit{SCI} provides a robust framework to capture the impact of this parameter on disease dynamics. A high clearance rate for needles implies a situation in which infected needles rapidly transition from an infected to an uninfected state. This scenario might arise in environments characterized by rapid viral decay in needles or where infected needles are promptly exchanged for non-infected ones, as observed in specific needle exchange programs \cite{davis2017needle, levengood2021supervised}. In this study, we assume that the clearance rate for needles in community 1 exceeds that of community 2. This discrepancy could occur, for example, if only community 1 participates in needle exchange programs. Specifically, we employ parameter values of $\epsilon_1=2$ and $\epsilon_2=0.33$ to represent this distinction. 

Despite this variation, mixing, characterized by inter-transmission between communities, plays a crucial role. Mixing tends to bring the disease dynamics of both communities closer to that of the community experiencing a lower clearance rate for needles ($\epsilon$). (\cref{fig:HCVResSummary} (a) and (d)). This convergence underscores the complex interplay between epidemiological parameters and social determinants within PWID communities. The \textit{SCI} offers valuable insights into infectious disease spread mechanisms (see \cref{tab:hcvSCI}). More specifically, it shows the community that has the largest influence on the spread of the disease (in this example, it is community 2, or the community with a lower $\epsilon$).

The two left subplots ((a) and (d)) in \cref{fig:HCVResSummary} demonstrate the disease dynamics for communities 1 (higher $\epsilon$) and 2 by simulating the $I_{E_1}$ and $I_{E_2}$ population sizes over time for four mixing scenarios. If there is no-mixing of these communities (isolated), we expect community 1 to have comparatively slower dynamics with lower infected populations, as shown by the blue curves in the plots in \cref{fig:HCVResSummary}. Additionally, we observe that even relatively small proportions of mixing of these two communities have a greater impact on the dynamics of community 1, and it tends to move towards dynamics similar to the community with a lower clearance rate of needles (i.e., community 2). \cref{tab:hcvSCI} provides the computational values for the basic reproduction numbers and the \textit{SCI}, and shows that the \textit{SCI} for community 2 is relatively higher than community 1. This indicates that community 2 is driving the dynamics of the infected population. Hence, if there is mixing, the population dynamics of community 1 will tend toward that of community 2.

We learn the importance of inequality: even with a higher clearance rate of needles in one community or the implementation of needle exchange programs, if the other community does not also have access to similar programs, then the shared use of needles undermines broader goals for a broader reduction in \gls{hcv} infections.  


\begin{table}
\centering
\begin{tabular}{|p{2cm}||p{0.95cm}|p{0.8cm}|p{1.65cm}|p{1.6cm}|p{1.65cm}|p{1.6cm}|}
\toprule
\rowcolor{gray!80}
Controlled parameter &community mixing & $\mathcal{R}_0^{\text{HCV(SD)}}$& $\mathcal{R}_{0_2}$: \textit{Perfect intervention} for community 1 & $\mathcal{C}_1$: $\frac{\mathcal{R}_0^{\text{HCV(SD)}}-\mathcal{R}_{0_2}}{\mathcal{R}_0^{\text{HCV(SD)}}}$ & $\mathcal{R}_{0_1}$: \textit{Perfect intervention} for community 2 &$\mathcal{C}_2$: $\frac{\mathcal{R}_0^{\text{HCV(SD)}}-\mathcal{R}_{0_1}}{\mathcal{R}_0^{\text{HCV(SD)}}}$ \\
\noalign{\vskip 0.2cm}
\hline
\multirow{4}{2.3cm}{$\epsilon_1=2$ and $\epsilon_2=0.33$. See \cref{fig:HCVResSummary} (a) \& (d). } & $0\%$ &$2.9012$ &$2.9012$ &$0$ &$1.3630$ &$0.5302$ \\  \cline{2-7}  
& $1\%$  & $2.9015$& $2.9012$ &$0.0001$ & $1.3630$ &$0.5302$\\ \cline{2-7}
& $5\%$ & $2.9089$& $2.9012$ &$0.0026$ & $1.3630$ &$0.5314$\\ \cline{2-7}
  & $100\%$ & $4.5331$ & $2.9012$&$0.3600$&$1.3630$ & $0.6993$ \\ \hline
  \noalign{\vskip 0.2cm}
  \hline
\multirow{4}{2.3cm}{$\varphi_1=0.05$ and $\varphi_2=5\times10^{-8}$. See \cref{fig:HCVResSummary} (b) \& (e). } &$0\%$ &$1.9640$ &$1.9640$ &$0$ &$0.9951$ &$0.4933$ \\ \cline{2-7} 
&$1\%$  & $1.9642 $& $1.9640$ &$0.0001$ & $0.9951$ &$0.4934$\\ \cline{2-7}
&$5\%$  & $1.9698 $& $1.9640$ &$0.0030$ & $0.9951$ &$0.4948$\\ \cline{2-7}
  & $100\%$ & $3.1137 $ & $1.9640$&$0.3692$&$0.9951$ & $0.6804$ \\ \hline
  \noalign{\vskip 0.2cm}
\hline
\multirow{4}{2.3cm}{$\tau_1=0.04$ and $\tau_2=0.0001$. See \cref{fig:HCVResSummary} (c) \& (f). } & $0\%$ &$8.2224$ &$8.2224$ &$0$ &$0.9899$ &$0.8796$ \\ \cline{2-7} 
&$1\%$ & $8.2228$& $8.2224$ &$0.0001$ & $0.9899$ &$0.8796$\\ \cline{2-7}
&$5\%$  & $8.2333 $& $8.2224$ &$0.0013$ & $0.9899$ &$0.8798$\\ \cline{2-7}
 &$100\%$ & $11.7122 $ & $8.2224$&$0.2980$&$0.9899$ & $0.9155$ \\
 \hline
\bottomrule
\end{tabular}
\caption{HCV mathematical model: Computed basic reproduction numbers and \gls{sciAb} values under varying levels of mixing and selected control parameters. 
Since community 2 is more vulnerable (eg: due to a lower needle clearance rate, $\epsilon$, compared to community 1), the basic reproduction number for isolated community 2 is higher than that for community 1. Consequently, a \textit{perfect intervention} in community 2 has a significant impact on the overall epidemiological dynamics ($\mathcal{C}_2 > \mathcal{C}_1$). Even minimal inter-community transmission increases the influence of community 2’s dynamics on community 1.
}
\label{tab:hcvSCI}
\end{table}

\subsubsection{HCV(SD) model: Self-clearance rate ($\boldsymbol{\varphi}$) and treatment rate ($\boldsymbol{\tau}$) of the community}
A high self-clearance rate implies a scenario where infected individuals swiftly transition from an infected to a non-infected state due to effective immune responses. Conversely, a low self-clearance rate indicates a slower transition \cite{ge2009genetic,hajarizadeh2013epidemiology}. Social determinants can significantly influence this parameter \cite{harris2013hepatitis}, making it a crucial control parameter in the HCV(SD) model to investigate its effect on disease dynamics. Specifically, we employ a higher self-clearance rate ($\varphi_1=0.05$) for community 1 and a lower self-clearance rate ($\varphi_2=5\times10^{-8}$) for community 2. These parameter values are selected to ensure that isolated community 1 exhibits endemic conditions while isolated community 2 experiences an epidemic.

Similarly, the treatment rate denotes the speed at which infected individuals receive medical treatment. A high treatment rate signifies efficient healthcare access and prompt initiation of treatment upon infection detection, leading to faster recovery and reduced transmission. Conversely, a low treatment rate suggests delays in healthcare access or treatment initiation, prolonging the duration of infection and potentially increasing transmission opportunities. Therefore, we utilize the treatment rate as another control parameter in our HCV(SD) model to investigate its impact on disease dynamics. Specifically, we assign a higher treatment rate ($\tau_1=0.04$) for community 1 and a lower treatment rate ($\tau_2=0.0001$) for community 1. Similar to the self-clearance parameter, these parameter values are chosen to ensure that isolated community 1 exhibits endemic conditions while isolated community 1 experiences an epidemic.

Disease dynamics, especially the early-infected population, under varying control parameters including self-clearance rate and treatment rates, are depicted in the middle ((b) and (e)) and right panels ((c) and (f)) of \cref{fig:HCVResSummary}. It should be noted that, in both cases, the endemic dynamics of community 2 when isolated (as indicated by the blue curve in the top panel) transitions to an epidemic state when there is a low-mixing rate, where $I_{E_1}$ does not converge to zero. Therefore, mixing heavily influences disease dynamics. Furthermore, community 2 exerts a dominant effect on the disease dynamics of the full model. This influence can be quantified using the \gls{sciAb}, and the corresponding \textit{SCI} values related to these parameters, self-clearance rate and treatment rate, are presented in \cref{tab:hcvSCI}. Note that in both cases, the \textit{SCI} for community 2 ($\mathcal{C}_2$) dominates. Therefore, implementing interventions in community 1 alone may not effectively reduce disease dynamics. Relying solely on the basic reproduction number for the entire population may not provide a comprehensive understanding of the disease dynamics. The impact of social factors can be quantified through the \textit{SCI}, providing insights for informing intervention strategies.

\subsection{Applying the \textit{SCI} to agent-based modeling (ABM)}\setcurrentname{Applying the \textit{SCI} to agent-based modeling (ABM)}\label{sec:RestABM}
In real-world data analysis, achieving a \textit{perfect intervention} scenario or obtaining data under strict no-mixing conditions may not always be feasible. One approach is to utilize modeling techniques, such as \glsdesc{ABM}, to simulate this scenario based on available data. Alternatively, expert knowledge can be incorporated to estimate parameters for such a setting, or early incidence data can be directly used, assuming minimal mixing during the initial outbreak period. 

In this section, we demonstrate the application of the \gls{sciAb} in analyzing infectious disease data by using an \gls{ABM} to generate simulated outbreak data. A flexible \gls{ABM} framework can be used to simulate interactions between individuals in two distinct social communities, each characterized by varying contact rates, movement behaviors, and infection probabilities. The model promotes transparency and interpretability of the underlying parameters. In our implementation, the \gls{ABM} comprises 2,000 agents distributed across a continuous two-dimensional spatial environment. The agents are equally divided into two spatially segregated communities, with community~2 (modeled as more vulnerable to disease) occupying the left half of the space and community~1 (modeled as less vulnerable to disease) the right half. Agents are initialized at random positions within their respective regions. Each agent exists in one of four epidemiological states: Susceptible, Exposed, Infectious, or Recovered (S-E-I-R). While this example is not provided to model any particular infectious disease, the S-E-I-R structure has been used to model a wide number of epidemic scenarios, including SARS-CoV-2 \cite{he2020seir}. 

The chosen hyperparameters and distributions for the agent-based simulation of disease dynamics in this section reflect underlying social inequalities that influence transmission. Disease spread is shaped by heterogeneity in contact rates, infection probabilities, and the durations of both the exposed and infectious periods. Specifically, infection probabilities are assigned using beta distributions: agents in community 2 are drawn from $\text{Beta}(2,5)$, and those in community 1 from $\text{Beta}(1,5)$, indicating higher susceptibility among community 2 individuals. The infectious period is heterogeneously distributed, with community 2 following an exponential distribution with a mean of 10 time units, and community 1 with a mean of 4 time units. The mean infectious period is denoted by $1/\gamma_i$ for community $i$. Similarly, the latent period (i.e., the time spent in the exposed state before becoming infectious) is modeled using exponential distributions, with community 2 having a mean of 4 time units and community 1 a mean of 1 time unit, denoted by $1/\epsilon_i$ for community $i$. Contact rates are also community-specific, sampled from log-normal distributions: community 2 follows $\mathcal{LN}(5,\,0.4^2)$ and community 1 follows $\mathcal{LN}(0.5,\,0.3^2)$, where the parameters correspond to the mean and variance of the underlying normal distribution. These values capture differences in social behavior, contact intensity, or other structural factors.

Each simulation is conducted over 250 discrete time steps. During this period, agents move locally within the spatial boundaries of their respective communities, with the exception of a small proportion (denoted by $m$) of individuals from community 2 (considered more vulnerable), who are permitted to temporarily move into the region occupied by community 1 (considered less vulnerable). Notably, infectious agents in community 2 remain stationary. At the onset of the simulation, one exposed agent is introduced into each community. At each time step, infectious agents may expose susceptible neighbors located within a fixed radius, with the probability of transmission determined by their individual contact rate and infection probability. Exposed agents transition to the infectious state after completing their latent period and subsequently progress to the recovered state following their infectious period.

The analysis begins by treating the communities as independent sub-populations with no-mixing. This is followed by a scenario incorporating partial mixing, where a proportion $m$ of individuals from community 2 temporarily move into community 1 at each time step. Although this specific ABM is used for demonstration purposes, the methodology can be extended to any ABM or real-world data. The primary focus is on examining epidemic curves to assess the impact of different mixing scenarios on disease spread, allowing for a clear analysis of the manner in which social forces drive disease dynamics through the \gls{sciAb}.

\subsubsection{\gls{ABM}-simulated data: Analyzing the impact of social segregation on disease spread using \gls{sciAb}}
\begin{figure}[ht]
    \centering
        \begin{subfigure}{0.45\textwidth}
        \centering
       \begin{overpic}[width=1\linewidth]{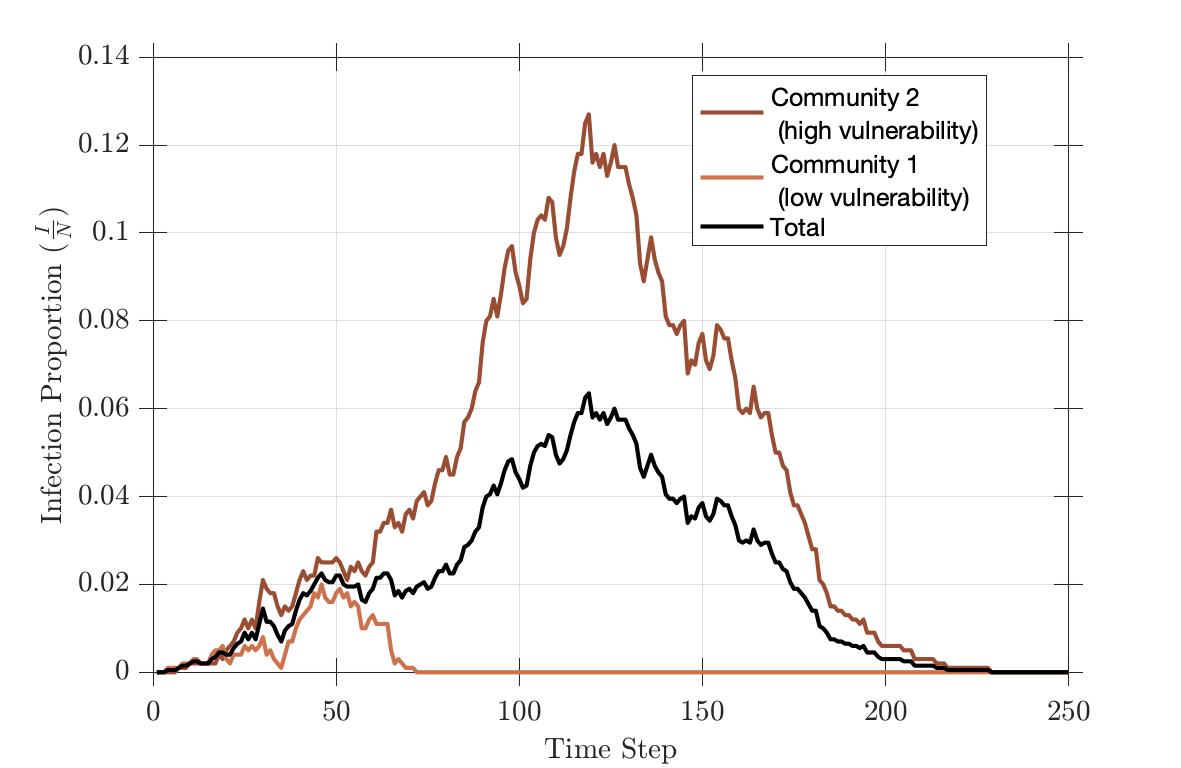}
        \put(-1,60){\textbf{(a)}}
    \end{overpic}
    \caption*{}
    \end{subfigure}
        \begin{subfigure}{0.45\textwidth}
        \centering
    \begin{overpic}[width=1\linewidth, valign=t]{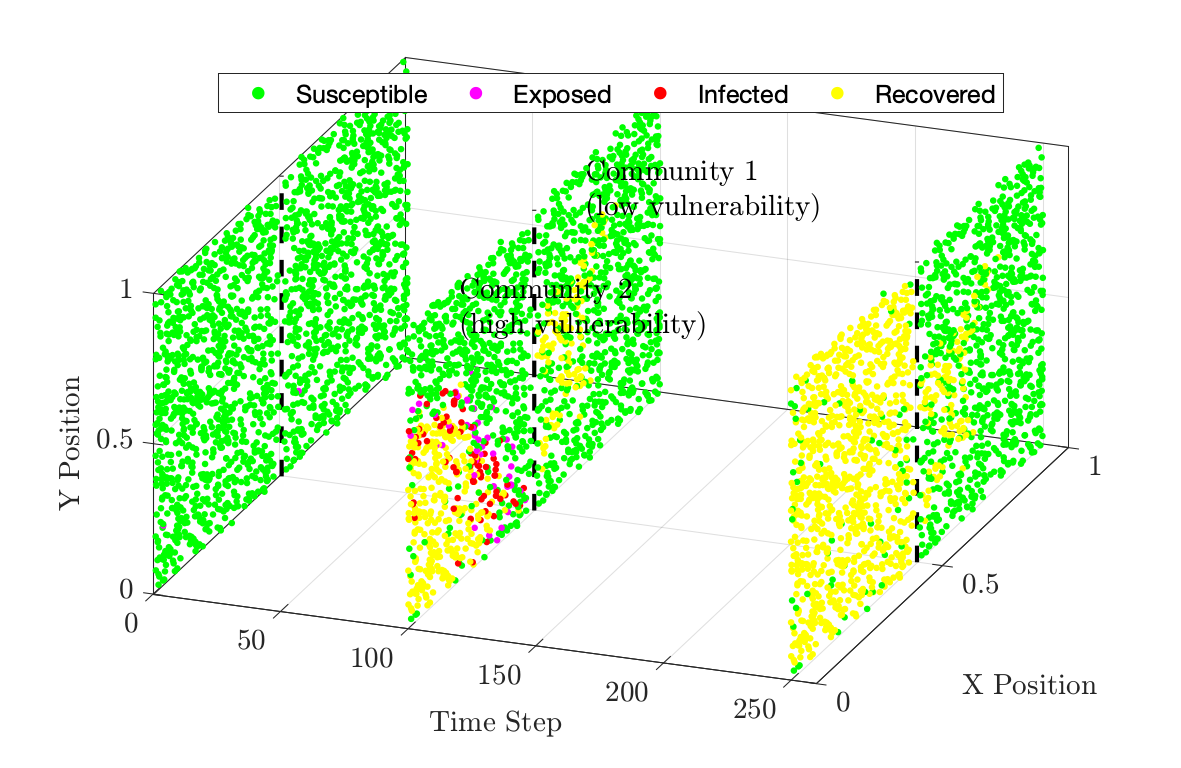}
        \put(5,60){\textbf{(b)}}
    \end{overpic}
    \caption*{}
    \end{subfigure}
    \caption{\textbf{No-mixing scenario dynamics in the agent-based model.} Visualization of the no-mixing scenario in a disease outbreak based on a single simulation. (a) illustrates the proportion of infectious individuals relative to the considered population size over time, depicting the dynamics of the infected population for community 1 (less vulnerable to disease, light-brown curve), community 2 (more vulnerable to disease, dark brown  curve), and the total population (black curve). (b) Three-dimensional visualization of epidemic states at selected time steps ($t=1$, $100$, and $250$). Each slice represents a 2D spatial distribution of agents in the population. Colors represent epidemic states: green for susceptible, magenta for exposed, red for infected, and yellow for recovered individuals. The dashed vertical lines indicate the midpoint dividing two social communities, labeled as community~1 (low vulnerability) and community~2 (high vulnerability).  In this case, the disease spreads only in the more vulnerable community (community 2), while the less vulnerable community (community 1) remains unaffected, despite the overall community appearing to have an epidemic.}
    \label{fig:NoMix1}
\end{figure}

\setcurrentname{\gls{ABM}-simulated data: Analyzing the impact of social segregation on disease spread using \gls{sciAb}}\label{sec:NomixABM}
As shown in \cref{fig:NoMix1}, when social communities do not interact during a disease outbreak, the overall system effectively decomposes into two independent sub-models, each governed by its own set of parameter values. However, if these underlying social disparities are not properly accounted for, public health authorities may mistakenly model the population as a single homogeneous entity. Such an oversight can result in ineffective prevention strategies at the population-level. This no-mixing scenario may emerge due to factors such as differences in awareness levels, structural barriers, or inherent social inequalities. In such cases, the disease burden may fall disproportionately on the more vulnerable community, while the less vulnerable community may remain largely unaffected. As a result, the aggregated population dynamics may misleadingly suggest an ongoing epidemic, even if only a subset of the population is significantly impacted. To simulate and analyze this scenario, we assume no inter-community movement by setting $m=0$.

From a data-driven perspective, empirical analyses do not inherently differentiate between models that incorporate social determinants and those that do not. Consequently, the estimated basic reproduction number for the full model is typically derived from parameter values-or infection data-that reflect aggregate population-level dynamics. In this context, the estimated basic reproduction number for the full population can be computed using the \gls{EG} method, as described in \cite{wallinga2007generation}, and is given by
\begin{equation*}
    \hat{\mathcal{R}}_0^{\text{SD}} = \left(1 + \frac{r}{\epsilon}\right) \left(1 + \frac{r}{\gamma}\right) = 2.2259.
\end{equation*}
where $r$, $1/\epsilon$, and $1/\gamma$ represent the estimated values of the initial infection growth rate, mean latent period, and mean infectious period of the total population, respectively. The computed value corresponds to the simulation outcome visualized in \cref{fig:NoMix1}.

Similarly, the basic reproduction numbers for each isolated community can be estimated. For the $i^{\text{th}}$ community, the basic reproduction number is given by:
\begin{equation*}
    \hat{\mathcal{R}}_{0_i} = \left(1 + \frac{r_i}{\epsilon_i}\right) \left(1 + \frac{r_i}{\gamma_i}\right),
\end{equation*}
where $r_i$, $1/\epsilon_i$, and $1/\gamma_i$ represent the estimated values of the initial infection growth rate, mean latent period, and mean infectious period for community $i$, respectively.

Accordingly, in a 2-community model, the \gls{sciAb} for community 2 can be estimated as follows (note that a similar formula can be derived for community 1 by interchanging the subscripts 1 and 2):  
\begin{equation}
    \hat{\mathcal{C}}_2=\frac{\left(1 + \frac{r}{\epsilon}\right) \left(1 + \frac{r}{\gamma}\right)-\left(1 + \frac{r_1}{\epsilon_1}\right) \left(1 + \frac{r_1}{\gamma_1}\right)}{\left(1 + \frac{r}{\epsilon}\right) \left(1 + \frac{r}{\gamma}\right)}.
\end{equation}  

The estimated basic reproduction numbers corresponding to the disease dynamics of communities 1 and 2 in the simulation visualized in \cref{fig:NoMix1} are given by $\hat{\mathcal{R}}_{0_1} = 1.2875$ and $\hat{\mathcal{R}}_{0_2} = 3.0362$, respectively. Consequently, the estimated \gls{sciAb} for communities 1 and 2 can be computed as:
\begin{align*}
    \hat{\mathcal{C}}_1 &\approx \frac{2.2259-3.0362}{2.2259}=-0.3640 \quad \text{and} \\ \hat{\mathcal{C}}_2 &\approx \frac{2.2259-1.2875}{2.2259}= 0.4216.
\end{align*}

In practical applications, \gls{sciAb} can be negative (as observed for $\hat{\mathcal{C}}_1$ in this case), indicating a counterintuitive effect on disease transmission when considering the total population. As illustrated in \cref{fig:NoMix1}, the black curve (representing infections in the full population) is a negatively shifted version of the dark brown curve (representing infections in a more vulnerable community). This arises because there are low infections in community 1 while the disease spreads within community 2. This implies that if one analyzes the entire population without accounting for social determinants, the estimated proportion of infected individuals appears lower due to the significantly reduced number of infections in community 1.

\subsubsection{\gls{ABM}-simulated data: Assessing the impact of community mixing on disease spread using \gls{sciAb}}

In this scenario, only the movement parameter $m$—representing the proportion of individuals from community 2 (more vulnerable) who temporarily move into community 1 (less vulnerable) at each time step—is varied, while all other hyperparameters and distributions remain identical to those in the no-mixing scenario. It is assumed that individuals in community 1 move only within their designated region and that infectious individuals remain stationary. Social community mixing is introduced by allowing a fraction of individuals from community 2 to temporarily move into the region of community 1 at each time step before returning to their original location. This scenario, particularly relevant for airborne pathogens, models workplace interactions where individuals from different social communities have the potential to come into contact. For demonstration purposes, the values of the chosen parameters emphasize the effect of social determinants, rendering community 1 less vulnerable to the disease. As a result, community 1 shows relatively few secondary infections from the initial case in the absence of mixing. The basic reproduction numbers at the community-level in the no-mixing scenario can be interpreted as the basic reproduction numbers under a \textit{perfect intervention}. For example, the basic reproduction number corresponding to a \textit{perfect intervention} targeting community 2 (the more vulnerable community) is equivalent to the basic reproduction number of community 1 under the no-mixing condition. Thus, the basic reproduction number for a \textit{perfect intervention} in community 2 is given by $\mathcal{R}_0^{\text{do}(\Gamma_2)} = \mathcal{R}_{0_1}$.

\begin{figure}[ht]
    \centering
    \begin{subfigure}{0.45\textwidth}
        \centering
\begin{overpic}[width=1\linewidth]{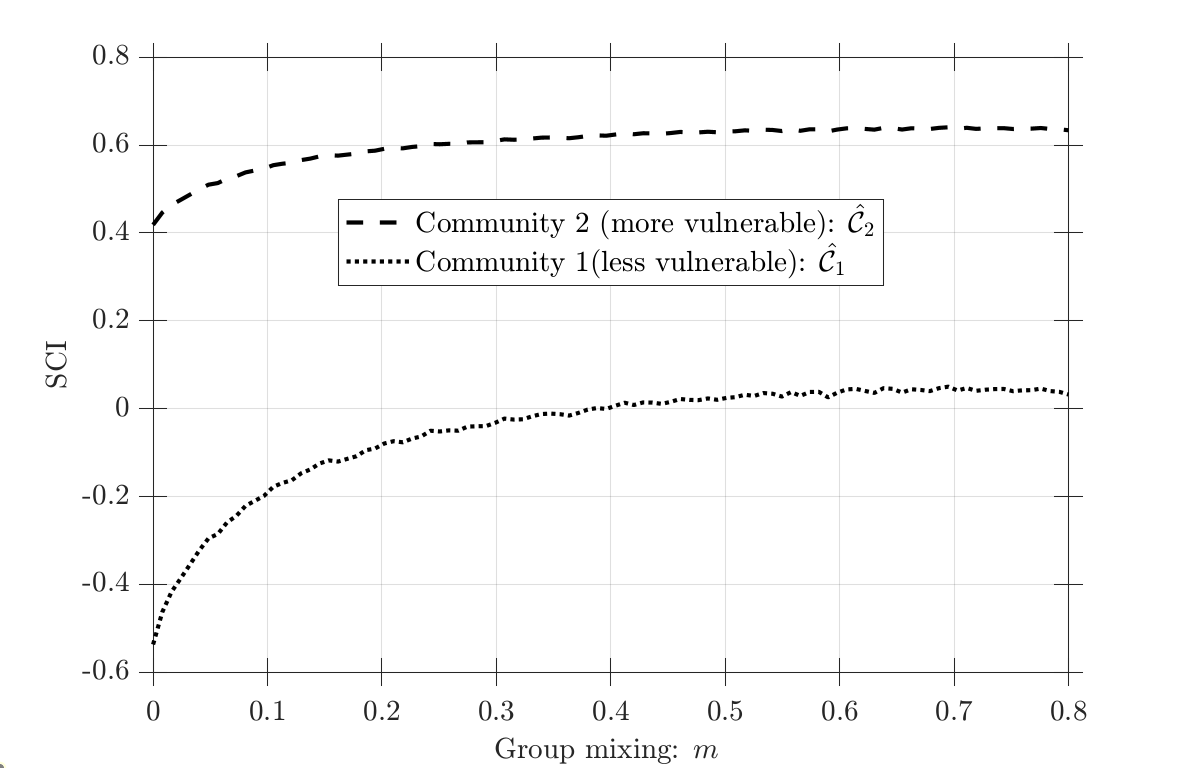}
\put(1,60){\textbf{(a)}}
\end{overpic}
        \caption*{} 
    \end{subfigure}
    \begin{subfigure}
        {0.45\textwidth}
        \centering
\begin{overpic}[width=\linewidth]{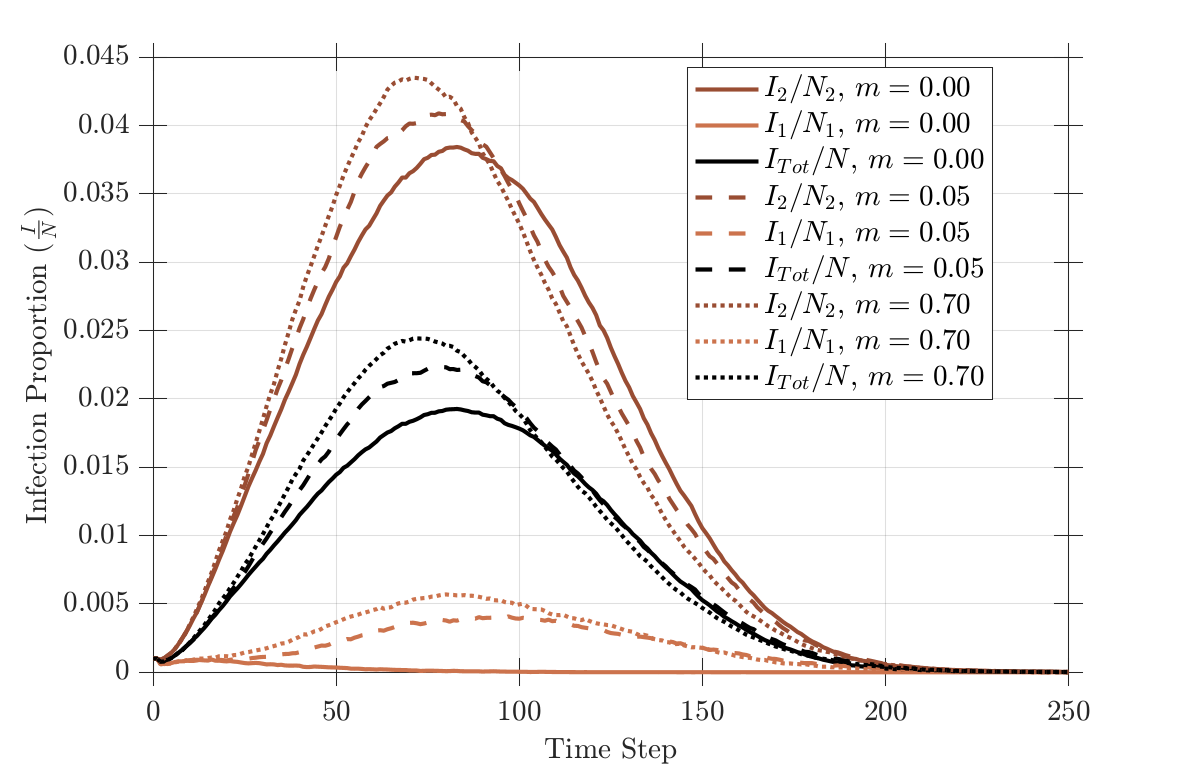}
\put(-3,60){\textbf{(b)}}
\end{overpic}
\caption*{}
    \end{subfigure}
    \caption{\textbf{Influence of community mixing on disease dynamics and \gls{sciAb} in an agent-based model.} (a) illustrates the variation of \gls{sciAb} for both communities as the mixing proportion \( m \) changes.  (b) illustrates the disease dynamics of community 1, community 2, and the total population for different values of the mixing parameter $m$. The results are based on the average of 1000 simulations. For $m = 0$, the disease in community 1 (the less vulnerable community) exhibits extinction or rapidly decays to zero following the initial infection. In scenarios with nonzero mixing, secondary growth of the infection in community 1 is observed.}\label{fig:abmSCIde}
    
\end{figure}

When using the \gls{ABM} approach, disease dynamics are analyzed not from a single simulation, but rather based on mean values across multiple simulation runs. Here, we conducted 1000 simulations and used the averaged incidence data to estimate the basic reproduction numbers and initial growth rates. The basic reproduction number for community 2 under the no-mixing condition is interpreted as representing a \textit{perfect intervention} in community 1. This value, estimated using the \gls{EG} method applied to the simulated data, is given by $\mathcal{R}_0^{\text{do}(\Gamma_1)} = \hat{\mathcal{R}}_{0_2} \approx 2.5488$. Similarly, the reproduction number for a \textit{perfect intervention} targeting community 2 is estimated as $\mathcal{R}_0^{\text{do}(\Gamma_2)} = \hat{\mathcal{R}}_{0_1} \approx 0.96456$. The growth rate of incidence in the total population, $r$, in the estimated \gls{sciAb} values depends on the extent of mixing between communities. To formally represent this community mixing, we define the parameter $m$ as the temporary movement proportion from community 2 to community 1. Since $r$ varies with the level of community mixing, it can be expressed as a function of $m$, denoted as $r(m)$. The estimated \gls{sciAb} values are then given by:  
\begin{align*}
    \hat{\mathcal{C}}_2(m) &\approx \frac{\hat{\mathcal{R}}_0^{\text{SD}}(m)-0.96456}{\hat{\mathcal{R}}_0^{\text{SD}}(m)} \quad \text{and} \\ \hat{\mathcal{C}}_1(m) &\approx \frac{\hat{\mathcal{R}}_0^{\text{SD}}(m) - 2.5488}{\hat{\mathcal{R}}_0^{\text{SD}}(m)},
\end{align*}
where $\hat{\mathcal{R}}_0^{\text{SD}}(m)=\left(1 + \frac{r(m)}{\epsilon}\right) \left(1 + \frac{r(m)}{\gamma}\right)$. Hence, for all mixing scenarios, we have $\hat{\mathcal{C}}_2 > \hat{\mathcal{C}}_1$ (see also \cref{fig:abmSCIde}(a)), which implies that the influence of community 2 (more vulnerable) on disease dynamics is consistently greater.  

\cref{fig:abmSCIde}(a) illustrates the behavior of \gls{sciAb} for both communities as the level of community mixing varies. When mixing is low, community 1 exerts a negative influence on the overall disease spread (\cref{fig:abmSCIde}(a) for $m \in [0, 0.4]$). In practical settings, the overall disease dynamics often reflect an average across both communities. However, such averaging can be misleading. For example, if one community undergoes an epidemic while the disease fades out in the other, the aggregated data may misleadingly suggest that both communities are experiencing an epidemic (\cref{fig:abmSCIde}(b)). In such cases, the \gls{sciAb} value for the community experiencing disease extinction will be negative (\cref{fig:abmSCIde}(a)), indicating a suppressive effect on the overall disease dynamics of the total population.

A negative \gls{sciAb} value reflects scenarios where the growth of an epidemic in the total population is suppressed due to the characteristics of its component sub-communities. As shown in \cref{fig:abmSCIde}(b), the epidemic growth rate in community~1 (the less vulnerable group) is low, while community~2 experiences a significantly higher rate of epidemic growth. This discrepancy can result in a total population epidemic growth rate that is lower than the rate in isolated community~2 when $m < 0.4$. Specifically, $\hat{\mathcal{C}}_1 < 0$ for $m < 0.4$ (see \cref{fig:abmSCIde}(a)), indicating that community~1  exerts a suppressive effect on the overall disease dynamics, when the mixing parameter is below $40\%$. The corresponding outbreak dynamics are presented in \cref{fig:abmSCIde}(b), which shows the infectious population fractions relative to their respective population sizes for various mixing levels: $m = 0$, $0.05$, and $0.7$. These curves are derived from the average of 1000 \gls{ABM} simulation runs. Notably, in \cref{fig:abmSCIde}(a), $\hat{\mathcal{C}}_1$ remains negative for $m = 0$ and $m = 0.05$, approaches zero around $m = 0.4$, and turns positive at $m = 0.4$.

As illustrated in \cref{fig:abmSCIde}(b), community~2 experiences an epidemic across all values of $m$, and the disease dynamics of the total population largely mirror those of community~2. In contrast, for $m = 0$, community~1 demonstrates rapid disease extinction following the initial outbreak. Under other mixing scenarios, community~1 exhibits slow, steady, and relatively small secondary growth. Consequently, because the epidemic dynamics of the total population are predominantly shaped by community~2, the overall curve is negatively influenced by the suppression in community~1. 

As depicted in \cref{fig:abmSCIde}(a), the \gls{sciAb} values of community~2 remain consistently positive, while those of community~1 are negative for $m < 0.4$. Hence, the infection dynamics of the total population are largely driven by community~2 (as also seen in \cref{fig:abmSCIde}(a)). Furthermore, the social factors of community~2 renders it both more susceptible to infection and more influential in driving disease spread. In contrast, community~1 is socially structured in a way that reduces both its risk of infection and its contribution to disease transmission. Additionally, as illustrated in \cref{fig:abmSCIde}(b), the infection curve of the total population may obscure important social inequalities in the outbreak. The proposed \gls{sciAb} metric captures such social inequalities and offers a more detailed perspective on the underlying dynamics, highlighting inequalities that conventional aggregate metrics may overlook.
\section{Discussion}

In this study, we apply the concept of causal inference through epidemiological models. Specifically, we utilize the concept of structural causal models with deterministic ordinary differential equations (ODEs) to address the question: ``What would happen if a \textit{perfect intervention} was applied to a given social community within epidemiological models considering the existence of the social inequalities?"

We follow the previous calls for more nuance in disease modeling \cite{powersmitchelltransmission} by introducing the \glsdesc{sci} to more rigorously examine how social forces craft the complexity of disease dynamics. To do this, we develop three case-examples. The first is a standard mathematical model, which is important for developing our approach using a clear and tractable example. We then develop a more complex mathematical model (\gls{hcv} transmission in a community of persons who inject drugs). In part due to its connection with addiction, \gls{hcv} has long been understood to be a disease where social determinants and access to resources affect the dynamics of the disease \cite{rourke2011social}. Lastly, we develop the \textit{SCI} using an entirely different computational approach: an epidemic setting, with an S-E-I-R structure, using an agent-based model. While our focus has been on the relationship between social forces and human infectious diseases,   we emphasize that our approach can apply to any epidemic setting where inequalities manifest, even in non-human pathogen systems, where select studies have identified analogous patterns in natural settings \cite{dwyer1997host}. In this sense, the \textit{SCI} is a relevant approach for the broader field of disease ecology. Furthermore, we hope to bring attention to related metrics such as the population attributable fraction (PAF) \cite{ml1953occurrence}.

\subsection{A metric for considering social forces in mathematical epidemiology: \textit{structural causal influence}}

The \gls{sciAb} emerges as a robust and detailed metric, as demonstrated in the example of homogeneous mixing with varying recovery rates for different communities (\cref{fig:HCVResSummary}). We argue that \gls{sciAb} can be used to understand the influence of each community on the spread of the disease, and therefore, can be deployed to find strategies for interventions. Specifically, the \gls{sciAb} offers several important features: 

\begin{enumerate}
    \item \textbf{Effectiveness of community-level interventions:} The \gls{sciAb} measures the impact of community-level \textit{perfect interventions}, allowing for the evaluation of the effectiveness of the intervention on a larger scale. This feature enables public health agencies to implement targeted strategies to potentially curb disease spread more efficiently.
    \item \textbf{Identification of influential communities:} The \gls{sciAb} can be used to identify the communities that have the most significant influence on the spread of a disease. By recognizing these key communities, public health practices can be more focused and strategic, optimizing resource allocation and intervention efforts.
    \item \textbf{Analysis of control parameters:} The \gls{sciAb} is capable of identifying the effects of various control parameters and determining which community-level parameters must be controlled. This insight is crucial for designing effective public health policies and interventions, ensuring that control measures are effective and efficient in mitigating disease spread.
\end{enumerate}

With these features in mind, we can consider two public health scenarios in which the \gls{sciAb} may find relevance: 

\begin{itemize}
    \item In the scenario involving communities with equal population sizes and uniform transmission rates, where one community (e.g., community 2) exclusively undergoes intra-community transmission. The disease is more likely to be transmitted to community 2 in most instances. If the mean infectious period of community 2 is higher than that of community 1 ($a < 0$), the influence of community 2 on disease dynamics is relatively greater than that of community 1 ( See Supporting Information Figure~S8).
    \item In the scenario involving communities with equal populations and uniform transmission rate, where there is no intra-community transmission for community 2: The mean infectious period of community 1 significantly influences the spread of the disease. If the mean infectious period for community 1 is higher than that of community 2 ($a > 0$), there is a higher probability that the basic reproduction number for the SIR(SD) model exceeds that of the classical SIR model (See Supporting Information Figure~S8).
\end{itemize}

\subsubsection{Interpreting \textit{SCI} values}

The \textit{SCI} is a relative metric, in other words, it quantifies the relative contribution of specific social communities to overall disease transmission in the population as a whole (see also Sections: ``\nameref{sec:mesSDOH}" and ``Agent-Based Model Incorporating Health Inequalities'' in the Supporting Information). Therefore, any interpretations of $\mathcal{C}_i$ can only be made with respect to other communities in the same population. In a population where the $\mathcal{C}_i$ of all communities is positive (i.e. $0 < \mathcal{C}_i \le 1$), targeting interventions towards the community with the highest $\mathcal{C}_i$ might be most effective for reducing infections in the overall population, but targeting interventions towards other communities might also reduce infections in the overall population. Conversely, if there is a population where at least one community has a negative $\mathcal{C}_i$ (i.e., $-1 < \mathcal{C}_i < 0$), targeting interventions to the community with a negative $\mathcal{C}_i$ instead of the other communities with positive $\mathcal{C}_i$ could be counterintuitive and might even increase infections in the overall population.


\subsection{Speculation and future directions}

The structural causal model framework can be further expanded by incorporating data-driven methodologies, such as Bayesian analysis (see \cite{pearl2010introduction, toth2022active}), to enhance causal inference in the context of social forces and infectious diseases. Additionally, another philosophical approach related to causation, termed Granger causality, can be applied to time series data within this field, providing a more sophisticated modeling framework \cite{shojaie2022granger}. 

Furthermore, we must explore the scalability of our approach. Our study used susceptible populations divided into two sub-communities. One should ask: What about the case where the number of social communities in epidemiological models is further increased (to more than two)?  In this case, transmission patterns, both between and within communities ($\beta_{ij}$'s), can be modeled using the transmission matrix presented in the ``Measures for quantifying the influence of  social inequalities in the SIR(SD) model" section of Supporting Information. This approach simplifies the epidemiological model with social forces employing matrix notation, which facilitates the calculation of the relevant basic reproduction numbers. Furthermore, as the number of communities increases, the relevant basic reproduction number can be estimated using simulated or public health data \cite{blumberg2013comparing, saucedo2019computing, blumberg2014detecting}. 

\subsection{Conclusions}

In summary, this study proposes theoretical ideas in computational epidemiology through the lens of the social forces that shape the dynamics of infectious diseases. Far from a strict political endeavor, our effort aims to generate tools that allow us to understand disease dynamics as they occur in the world as it exists, rather than as idealized abstractions naive to the reality of how societies are structured. \vskip6pt

\section*{Acknowledgements}
The authors would like to thank K. Kabengele and members of the OgPlexus for comments on a draft of the manuscript. In addition, the authors thank the organizers and participants in the following scientific meetings: ``Yale Inference Workshop: Probing the Nature of Inference from Data Models and Simulations across Disciplines" (December 2023);  ``Statistical Methodologies for Mitigating Disparities in Medicine" panel at the New England Statistical Society Annual Spring Meeting (June 2023); and the ``Modeling and Theory in Population Biology" meeting at the Banff International Research Station (May 2024). Ideas related to this manuscript were discussed at these various gatherings. 

\section*{Funding}
This work was supported by the A*STAR National Science Scholarship, Singapore (S.N.M.), the Seesel Postdoctoral Fellowship from Yale University (S.S.), the Robert Wood Johnson Ideas for an Equitable Future Award (S.S. and C.B.O), and the Mynoon Doro and Stephen Doro MD, PhD Family Private Foundation Fund (S.S. and C.B.O.).

\section*{Competing interests}
The authors declare no conflicts of interest.

\section*{Author contributions} Conceptualization: SS and CBO. Model development: SS and CBO. Visualization: SS and SNM. Analysis: SS and CBO. Interpretation: SS, SNM, SVS, LC, and CBO. Writing—original draft: SS, SNM, and CBO, Writing—review and editing: SS, SNM, SVS, LC, and CBO. Supervision: SVS, LC, CBO. Funding aquisition: CBO. All authors gave final approval for publication and agreed to be held accountable for the work performed therein. 

\section*{Data Availability} 
Data can be found within the manuscript or within cited references. Code is available at \href{https://github.com/OgPlexus/SCI1}{github.com /OgPlexus/SCI1}





\clearpage
{\small \bibliographystyle{mystyle}  \bibliography{refs}}
\clearpage
\counterwithin*{section}{part}
\renewcommand{\thesection}{S\arabic{section}}

\setcounter{figure}{0}
\renewcommand{\thefigure}{S\arabic{figure}}
\setcounter{table}{0}
\renewcommand{\thetable}{S\arabic{table}}



\end{document}

%% file: glossary.tex
\newabbreviation{ode}{ODE}{ordinary differential equation}

\newglossaryentry{SDOH}{
    name={SDOH},
    description={social determinants of health},
    first={social determinants of health (SDOH)},
    short={SDOH}
}

\newabbreviation{dfe}{DFE}{disease-free equilibrium}
\newabbreviation{mfe}{MFE}{mutant-free equilibrium}
\newabbreviation{covid}{SARS-CoV-2}{severe acute respiratory syndrome coronavirus 2}
\newabbreviation{hcv}{HCV}{hepatitis C virus}

\glsxtrnewsymbol[description={basic reproduction number}]{r0}{\ensuremath{\operatorname{\mathcal{R}_0}}}

\glsxtrnewsymbol[description={relative reproduction number}]{rrn}{\ensuremath{\operatorname{\mathcal{A}}}, RRN}

\glsxtrnewsymbol[description={relative force of infection}]{rfoi}{\ensuremath{\operatorname{\mathcal{F}_i}}, RFoI}

\glsxtrnewsymbol[description={structural causal influence}]{sci}{\ensuremath{\operatorname{\mathcal{C}_i}}, SCI}

\glsxtrnewsymbol[description={basic reproduction number for the compartmental model incorporating social determinants of health (SDOH)}]{r0SD}{\ensuremath{\operatorname{\mathcal{R}_0^{\text{SD}}}}}

\glsxtrnewsymbol[description={\textit{“perfect” interventions}}]{do}{\ensuremath{\operatorname{do(\Gamma)}}}

\newabbreviation{sciAb}{SCI}{structural causal influence}
\newglossaryentry{ABM}{
    name={ABM},
    description={agent based modeling},
    first={agent based modeling (ABM)},
    short={ABM}
}

\newglossaryentry{EG}{
    name={EG},
    description={exponential growth},
    first={exponential growth (EG)},
    short={EG}
}